  \documentclass[fleqn,11pt,a4paper]{article}
  \usepackage[dvips]{graphicx}
 \def\Dot #1#2#3#4{%
 \begin{picture}(40.00,40.00)
 \unitlength=0.40mm
 \special{em:linewidth 0.4pt}
 \linethickness{0.4pt}
 \put(20.00,20.00){\makebox(0,0)[cc]{${#2}$}}
 \put(30.00,10.00){\makebox(0,0)[cc]{${#1}$}}
 \put(20.00,0.00){\makebox(0,0)[cc]{${#4}$}}
 \put(10.00,10.00){\makebox(0,0)[cc]{${#3}$}}
 \end{picture}}

\begin{document}

\title{   Magnetic dot arrays modeling via the system of the radial
          basis function networks      }

\author{\small      Denis Horv\'ath $^{a}$,
                    Martin Gmitra $^{a}$
                    and Ivo V\'avra$^{b}$                    \\
        $^a$ Department of Theoretical Physics and Geophysics,\\
             University of P.J.\v{S}af\'arik, \\
             Moyzesova 16, 040 01 Ko\v{s}ice, \\
             Slovak Republic \\
        $^b$ Institute of Electrical Engineering SAS, \\
             D\'ubravsk\'a cesta 9, 84239 Bratislava, \\
             Slovak Republic  }

 \date{}
 \maketitle

 \begin{abstract}
  Two dimensional square lattice general model of the magnetic dot
  array is introduced. In this model the intradot self-energy
  is predicted via the neural network and interdot magnetostatic
  coupling is approximated by the collection of several
  dipolar terms. The model has been applied to disk-shaped
  cluster involving $193$ ultrathin dots and $772$ interaction centers.
  In this case among the intradot magnetic structures
  retrieved by neural networks
  the important role play single-vortex magnetization modes. Several
  aspects of the model have been understood numerically
  by means of the simulated annealing method.
 \end{abstract}

\section{Introduction}

 In the recent years there is a remarkable progress
 in the technology of the nanofabrication of well defined magnetic
 materials. The material nanoscience based on the epitaxial
 and lithographic techniques \cite{Chou97} allows the fabrication
 of the regular arrays of the magnetic particles-dots
 of well controlled and interesting shape \cite{TWu99}, lattice geometry and composition.
 The increasing technological flexibility calls for further physical ideas,
 which should be incorporated into design of the artificial nanoscale
 magnetic systems.

 The uniformity of the polarization is the general basic aspect
 discussed in the connection with the small magnetic particles. The
 concept of uniformly polarized particle is justified only
 for the particles of an intermediate size
 \cite{Aharoni91,Aharoni91nov}. In the theory \cite{Gushlienko99}
 the magnetostatic coupling was derived for the homogeneously
 polarized and saturated cylindrical dots on a rectangular lattice.
 More restrictive are conditions of the simulation \cite{Camley98},
 where each dot of array is substituted by a single dipolar moment. This
 approximation can be used only for monodomain dots separated
 by a sufficiently large distances. When a dot array is represented
 by a system of the interacting dipoles, the search for the ground
 state configuration leads to the formulation typical for
 the classical dipolar lattices \cite{Tisza46,Brankov87,Prakash90}.
 The violation of the intradot homogeneity stems from the competition
 between the magnetostatic, anisotropy and exchange energy terms.
 The analytical model of the dot array going towards
 the non-uniformity was proposed in \cite{GushSM14}. In this model
 the interactions of dots were described by the quadrupolar terms.

 The problem of the calculation of the magnetization field of a dot array
 can be in principal formulated in the terms of classical micromagnetic
 theory \cite{Brown62}. Due to complexity of the problem, the important
 role in its treatment will play the numerical simulations. They require
 the implementation of the sufficiently dense discretization within
 the each ferromagnetic dot. As usual, the magnetic part of the system
 can be subdivided into interacting dipoles
 or grains \cite{Berkov97,Miles91}, small ferromagnetic cubes \cite{Bertram88}
 or finite elements \cite{Schrefl99}. Then the optimum spacing
 of the mesh nodes is determined by a minimum magnetic length scale
 (exchange, wall) of the system. For the majority of ferromagnetic
 materials, the comprehensive micromagnetic description is attained when
 the size of discretization elements decreases into the nanometer
 regime \cite{Schrefl99}. Thus, the simulation of a single ultrathin
 dot of the micrometer size requires about $10^6$ nodes, although,
 the qualitative simulations can be realized even for $10^2-10^4$
 nodes \cite{Popkov99,Labbe99}. From this we can conclude 
 that both detailed and
 truncated micromagnetic description of many-dot array represents
 rather demanding computational task. We summarize, that principal
 difficulties of micromagnetic dot array analysis come from: (i)~the
 interplay of the phenomena on the intradot (exchange, domain wall)
 and external geometric length scales; (ii)~complexity of
 the magnetic structure of the non-uniformly polarized dots;
 (iii)~long-range magnetostatic interdot interactions.

 To make the problem of the magnetic ordering of dot arrays
 tractable by a moderate computer facilities, we have developed
 method, which works on a much coarser mesh than usual
 discretization schemes allow (except the adaptive and multigrid methods).
 Its general idea is the simultaneous simulation of the intradot -
 micromagnetic and multidot scales. This idea was strongly inspired by the
 multiscale approach \cite{Abraham98}. At the present stage
 of the project, the multidot part of simulation has been
 developed separately and the behaviour of the small-intradot
 scales has been treated only phenomenologically. The approach allows
 a remarkable increase of the simulation speed, indeed, the price payed
 to the scale separation is the appearance of additional parameters.
 The completing of the project needs support of the algorithms
 of the parameter estimation developed on the basis of
 the standard micromagnetic simulations. Let us to note
 that similar problems were solved in a cellular automaton
 version of the molecular dynamics \cite{Lejeune99}.

 In this paper we presented the results of the simulation
 of magnetic properties of quasi-two-dimensional cluster
 of ultra-thin magnetic dots on the square lattice.
 The phenomenological aspect of our model is a variable
 intradot inhomogeneity. The formalism we developed
 for this aim has been adopted from the models of the
 neural networks.

 From the point of view of the information theory,
 the neural networks are continuous, unique mappings
 constructed from the system of known activation
 functions. The synaptic weights of these activation
 functions are adjusted by the training process.
 The standard problem, which can be effectively
 solved by the neural networks is the association
 of the input patterns (in our case inputs are
 effective magnetic moments) with the desired outputs
 (magnetic self-energy of dot). There are many
 applications, where neural networks can be implemented.
 They allow interpolation of the data generated by the simulations
 or experiment. The example of the physical application
 is \cite{BrownGibbs96}, where neural network was used to fit
 a complicated analytic potential to the set of ab initio data.
 In \cite{LiebowitchL94}, the Hopfield type of the
 interaction matrix was suggested to simulate
 the dynamics of the complex protein molecule. The specific
 magnetic application represents the solution
 of the magnetic inverse problem \cite{Jones99}.

 For the purpose to model the variable magnetic intradot
 inhomogeneity we adopted the theory of the radial basis
 function networks (RBFN) \cite{Haykin98}. The RBFN variant
 of the neural network was chosen, because its ingredient
 is a straightforward and explicit estimation of the
 synaptic weights, which allows more transparent analysis
 of the physical symmetries.

 The aim of the paper is to make general presentation of the model and present some numerical results.
 The paper is organized as follows: In Section~\ref{Modell}
 the model of the dot array energy functional is introduced.
 This Section consists of two parts: in Subsection~\ref{Model0} 
 we introduced the general phenomenological concept of 
 the nonuniform magnetization, which utilizes self-energy 
 interpolation by means of RBFN approach.
 In Subsection~\ref{Model1} the interaction between dots
 is introduced. In Section~\ref{Square} our method
 is applied to the ultrathin square dots, where tendency
 of the formation of the vortex intradot phase prevails.
 Section~\ref{Simull} provides some details about
 the implementation of the simulated annealing
 algorithm to the problem of the total 
 energy minimization.
 Finally, in Section~\ref{NumSimull} 
 we bring examples of 
 the numerical simulations.

 \section{Model}\label{Modell}

 \subsection{Intradot self-energy}\label{Model0}

 The microstate of the system of $N$ magnetic dots
 is described by $N_{\rm c}-by-N$ effective magnetic
 moments
 \begin{equation}
 {\bf m}_{in}\,,
 \qquad  i=1,2,\ldots, N\,,
 \qquad  n=1,2,\ldots, N_{\rm c}\,,
 \label{moments0}
 \end{equation}
 which are associated with the magnetization field of dot
 ${\bf M}(x,y)$ via the volume averages
 \begin{equation}
 {\bf m}_{in}= m_{\rm sat}^{-1} \,\,\,
 \int_{ (x,y) \in \triangle_{in} }\,
 {\rm d}x \,{\rm d}y\,{\bf M}(x,y)\,\,.
 \label{eqqq1}
 \end{equation}
 The formal integration
 is performed here over $n$th
 volume element
 of $i$th dot
 labeled as
 ${\triangle}_{in}$.
 The saturation magnetic moment
 $m_{\rm sat}=I_{\rm s} V_{\rm d}/N_{\rm c}$,
 where $I_{\rm s}$ is the saturation magnetization
 of the dot, plays role of the normalization factor
 in Eq.(\ref{eqqq1}). Thus for the effective
 magnetic moments we have the bounding
 \begin{equation}
 m_{in,x}^2+m_{in,y}^2\leq 1\,\,,
 \label{bounds}
 \end{equation}
 where $x,$ $y$ subscripts refer to the Cartesian components
 of the effective magnetic moments ${\bf m}_{in}= m_{in,x}
 {\bf e}_x + m_{in,y} {\bf e}_y$, where 
 ${\bf e}_x$ and ${\bf e}_y$
 are the Cartesian unit vectors. The "softness"
 of ${\bf m}_{in}$ expressed by Eq.(\ref{bounds})
 is the important model aspect, which differs from the
 fundamental Brown's postulate \cite{Brown62}.
 The reason for this modification 
 is that our model is formulated for sufficiently
 larger elements than classical 
 micromagnetic approach.
 The total magnetization per dot per 
 interaction center is given by
 \begin{equation}
 {\bf m}=
 \frac{1}{N\,N_{\rm c}}\,
 \sum_{i=1}^N\,
 \sum_{n=1}^{N_{\rm c}}\,
 {\bf m}_{in}\,\,\,.
 \label{magnetiz}
 \end{equation}
 The effective moments characterizing $i$th dot inside $N$-cluster
 are distributed around the dot center ${\bf R}_i$ and located
 at $N_{\rm c}$ positions
 \begin{equation}
 {\bf X}_{in}= {\bf R}_i+{\bf r}_n\,,
 \qquad
 n=1,2,\ldots, N_{\rm c}\,,
 \label{intcenter}
 \end{equation}
 where ${\bf r}_n$
 are some
 relative coordinates of the interaction
 centers. From the assumption that identical dots
 are arranged into array
 it follows that system
 of ${\bf r}_n$
 vectors is independent
 of the dot position
 inside the cluster.

 In a quasi-two dimensional systems, where ${\bf m}_{in}$ is
 confined to $x-y$ plane, reduced
 information about $i$th
 dot microstate
 is involved in $2 N_{\rm c}$ dimensional
 row vector
 \begin{eqnarray}
 {\tilde m}_i &\equiv &
 \left[\,
 {\bf m}_{i 1}\,,\,
 {\bf m}_{i 2}\,,\,
 \ldots\,,\,
 {\bf m}_{i N_{\rm c}}
 \,\right]\,
 \label{moments}
 \\
 & = &
 \left[\,
\underbrace{m_{i1,x}, m_{i 1,y}}_{{\rm center}\,\, {\bf X}_{i 1}},
\underbrace{m_{i2,x}, m_{i 2,y}}_{{\rm center}\,\, {\bf X}_{i 2}},
\ldots\,,
\underbrace{m_{i N_{\rm c},x}, m_{i N_{\rm c},y}}_{{\rm center}\,\,
{\bf X}_{i N_{\rm c}}}\,
\right]\,.
 \label{moments01}
\end{eqnarray}

 We continue with the construction of relations associating
 the effective dot moments from Eq.(\ref{moments})
 with the corresponding self-energies.
 Here the intradot self-energy is understood as a part
 of the total energy,
 which includes only anisotropy, exchange and intradot
 magnetostatic energy contributions. The Zeeman term
 and interdot magnetostatic terms, which do not contribute
 to the self-energy are defined independently
 of the neural network part of the model.

 The construction of the self-energy formula
 $E^{\rm self}({\tilde m}_i)$ is based on a proper
 choice of the set of special $2\,N_{\rm c}$-dimensional
 memorized vectors (the input patterns of neural networks)
 of the type Eq.(\ref{moments01}). The memorized vectors
 are constructed by putting
 into row $N_{\rm c}$ two-dimensional
 vectors ${\bf p}_n^{(q)}$
 \begin{equation}
 {\tilde p}^{(q)}= \left[
 \,p_{1,x}^{(q)},  p_{1,y}^{(q)},
   p_{2,x}^{(q)},  p_{2,y}^{(q)},
  \ldots\,,
   p_{{N_{\rm c}},x}^{(q)},
   p_{{N_{\rm c}},y}^{(q)}
  \,
  \right]\,.
  \label{moments1}
  \end{equation}
 Here, the superscript $q$ identifies so called feature \cite{Kohonen95}.
 In our case it is an integer from the set $\Lambda_Q \equiv
 \{0,1,\ldots,Q-1\}$. In the analogy with Eq.(\ref{moments}),
 the subscripts of $p_{n,x}^{(q)},p_{n,y}^{(q)}$
 run over the interaction centers
 ${\bf r}_n$.

 To measure the differences between the configurations
 we have introduced the Euclidean norm
 \begin{equation}
 ||\,{\tilde m}\,||= {\tilde m} \cdot {\tilde m}^{\rm T}
 \label{metrix}
 \end{equation}
 written here for some magnetic moment ${\tilde m}$
 [again encoded via the rule from Eq.(\ref{moments01})].
 The superscript ${\rm T}$ from Eq.(\ref{metrix})
 denotes the vector transposition.

 We start the construction of $E^{\rm self}({\tilde m}_i)$
 by assuming that self-energy is known for $Q$ memorized
 vectors
 \begin{equation}
 E^{\rm self}({\tilde p}^{(q)})=
 {w}^{(q)}\,,
 \qquad q\in \Lambda_Q\,,
 \label{condition1}
 \end{equation}
 where $w^{(q)}$ are free parameters
 of our model. The quality of the interpolation
 via RBFN depends on the choice of the basis
 functions and corresponding weights.
 Most convenient for our preliminary purposes
 seems to be the use of Nadaraya-Watson regression
 estimator \cite{Haykin98}. According to this,
 the self-energy input-output relation can be written
 \begin{equation}
 E^{\rm self}({\tilde m})=
 \displaystyle{\sum_{\scriptstyle q=0}^{Q-1}}\,
 {w}^{(q)}\,
 \psi^{(q)}({\tilde m})\,,
 \label{EqNW1}
 \end{equation}
 where
 \begin{equation}
 \psi^{(q)}({\tilde m})
 =\frac{
 \displaystyle{
 \exp
 \left(
 \,-\frac{Q}{d_{\max}^2}
 \,||{\tilde p}^{(q)}-{\tilde m}||^2
 \,\right)}}{
 \displaystyle{
 \sum_{\scriptstyle q=0}^{Q-1}}\,
 \exp\left(
 \,
 -\frac{Q}{d_{\max}^2}
 \,||{\tilde p}^{(q)}-{\tilde m}||^2
 \,\right)}
 \label{EqNW}
 \end{equation}
 are radial basis
 functions $ \{\psi^{(q)}({\tilde m})|$,
 $  q\in\Lambda_Q\}$ satisfying
 the normalization conditions
 \begin{equation}
 \sum_{q=0}^{Q-1}\,
 {\psi^{(q)}}({\tilde m})=1\,,
 \qquad 0 \leq \psi^{(q)}({\tilde m})\leq 1\,.
 \end{equation}
 The dot index of ${\tilde m}$ was omitted whenever the distinctions
 between the individual dots is unimportant. The choice of the
 dispersion $d_{\rm max}/(\sqrt{2}\,Q)$ in Eq.(\ref{EqNW}),
 where
 \begin{eqnarray}
 d_{\max} &\equiv &
 \begin{array}{c}
 \\ \max \\
 q,q'\in \Lambda_Q
 \end{array}
 \,||\,{\tilde p}^{(q)}-{\tilde p}^{(q')}\,||
 \end{eqnarray}
 is consistent with the recommendation \cite{Haykin98}.
 Speaking in terms of the neural networks
 $\psi^{(q)}({\tilde m})$ is the activation function,
 which determines $q-$th neuron's response to a given
 input ${\tilde m}$ and self-energy parameter
 $w^{(q)}, q\in \Lambda_Q$ is
 the optimized weight of the link
 between the input and output layer of
 the network.

 Having established
 the formula for the calculation
 of the self-energy, it is easy
 to prove that if
 the systems of the memorized
 vectors is composed
 from the conjugate vector pairs
 ${\tilde p}^{(q)}$, ${\tilde p}^{(q')}$
 with the same self-energy parameters
 \begin{equation}
 {\tilde p}^{(q)}=-{\tilde p}^{(q')}\,,
 \qquad
 {w}^{(q)}={w}^{(q')}
 \label{symmetry}
 \end{equation}
 RBFN self-energy form
 including the typical combination of terms
 \begin{displaymath}
 w^{(q)} \Big(
 \exp(-({Q}/{d_{\max}^2})\, ||{\tilde p}^{(q)}-{\tilde m}||^2 )
 +\exp(-({Q}/{d_{\max}^2})\, ||{\tilde p}^{(q)}+{\tilde m}||^2 )
 \Big)
 \end{displaymath}
 posses the reflection symmetry
 \begin{equation}
 E^{\rm self}({\tilde m})=
 E^{\rm self}(-{\tilde m})\,\,.
 \label{invariance1}
 \end{equation}
 To analyze the configuration snapshots
 generated during the simulation process
 we have introduced so called
 feature map ${\cal F}$.
 It associates any pattern vector
 ${\tilde m}$ with the feature $q^{\ast}\in\Lambda_Q$
 \cite{Haykin98}. The feature
 $q^{\ast}$ identifies
 the index of a nearest memorized vector
 ${\tilde p}^{(q^\ast)}$:
 \begin{eqnarray}
 {\cal F}: \qquad
 q^{\ast}&=&{\cal F}({\tilde m})\,,
 \\
 ||\,{\tilde p}^{(q^{\ast})}-{\tilde m}\,||
 &=&
\begin{array}{c}
\\
\min \\
q\in \Lambda_Q
\end{array}\,||\,{\tilde p}^{(q)}-{\tilde m}\,||\,.
\label{attrS}
\end{eqnarray}
 The classification of the features
 performed for the whole
 cluster gives rise to the
 $N-$component vector ${\cal F}({\tilde m}_1),
 {\cal F}({\tilde m}_2),
 \ldots,
 {\cal F}({\tilde m}_{N})$.
 The information about
 this vector can be concentrated
 to the form of the sample averages
\begin{equation}
{\overline n}{(q)}=\frac{1}{N}
\sum_{i=1}^N
\delta_{q,{\cal F}({\tilde m}_i)}\,,
\label{concentn}
\end{equation}
where $\delta$ is the usual Kronecker symbol.

\subsection{Interdot interactions and interactions
            with the external field}\label{Model1}

\begin{figure}
\begin{center}
\caption{ The schematic view
          on the calculation
          of energy functional
          using RBFN approach:
a)~the magnetic state of the dot array represented by the system
   of the eight-dimensional vectors;
b)~RBFN applied as a predictor of the self-energy output
   of $i$th dot from the input vector ${\tilde m}_i$;
c)~the interaction of dots (RBFN)
   mediated by the effective magnetic moments.
}
\vspace{2mm}
\includegraphics[width=10.0cm,height=11.0cm]{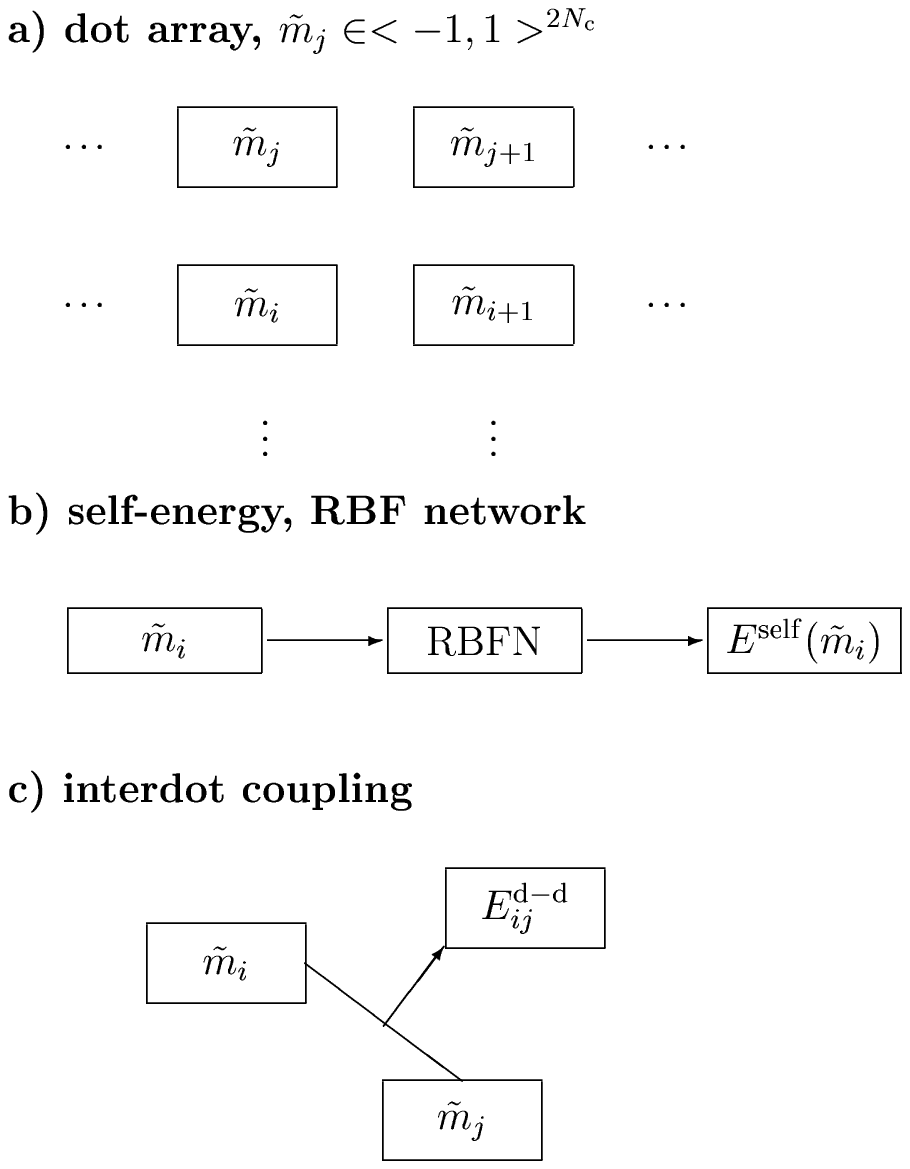}
\label{Fig3}
\end{center}
\end{figure}

The standard assumption about the interdot interactions is
that they are essentially magnetostatic \cite{Gushlienko99}.
By using the concept of the effective moments and interaction
centers one can construct easily the interdot interaction
potential. For this aim we expressed the energy
contribution $E^{\rm c-d}_{ik,j}$ consisting
of the dipole-dipole interactions
of the effective moment ${\bf m}_{in}$
[located at the $n$th center of $i$th
dot at the position ${\bf X}_{in}$]
with $N_{\rm c}$ effective moments of
$j$th dot ($j\neq i$)
\begin{equation}
 E^{\rm c-d}_{in,j}=
 \lambda\,\sum_{\alpha=x,y}\,
 m_{in,\alpha}\,
 \sum_{\beta=x,y}\,
 \sum_{s=1}^{N_{\rm c}}\,
 A_{injs,\alpha\beta}\,m_{js,\beta}\,\,\,,
 \label{dipolar}
 \end{equation}
 where
 \begin{eqnarray}
 A_{injs,\alpha\beta}&= &J_{\alpha\beta}\left(\,
\frac{{\bf X}_{in}-{\bf X}_{js}\,}{a}
 \right)\,,
 \qquad
 J_{\alpha\beta}({\bf x})=
 \frac{
 |{\bf x}|^2 \delta_{\alpha\beta}
 - 3 x_{\alpha}
x_{\beta}\,}{|{\bf x}|^{5}\,}\,.
\label{AQab}
\end{eqnarray}
Because ${\bf X}_{in}$, ${\bf X}_{js} $
are scaled by the lattice spacing $a$,
the energy dimension is absorbed
into dipolar constant
\begin{equation}
\lambda=\frac{m_{\rm sat}^2}{ 4\pi\mu_0\,\mu_{\rm r}\,a^3}\,,
\label{lambdaeq}
\end{equation}
where $\mu_{\rm r}$ is the relative permeability of the matrix.
For the interdot magnetostatic
energy of dot pair $(i,j)$ we obtain the expression
\begin{equation}
E^{\rm d-d}_{ij}=\sum_{n=1}^{N_{\rm c}}\,E^{c-d}_{in,j}\,\,.
\end{equation}
The key remaining contribution is the Zeeman energy.
For $i$th dot interacting with the
external magnetic field ${\bf H}$ we obtain
\begin{equation}
E^{\rm H}_i = - m_{\rm sat}\,{\bf H}
 \cdot \sum_{n=1}^{N_{\rm c}} {\bf m}_{in}\,.
\end{equation}
In further, to characterize the external field,
we have used the reduced undimensional
field ${\bf h}=m_{\rm sat}\,{\bf H}/\lambda$,
its components $h_x={\bf h} \cdot {\bf e}_x$ $=h \cos\theta$,
$h_y={\bf h}\cdot {\bf e}_y$$=h \sin\theta$ and polar angle
$\theta$. The final form of total energy functional
is then given by
\begin{equation}
 E=\sum_{i=1}^N
 \left\{
 E^{\rm self}({\tilde m}_i)
 + E^{\rm H}_i +
 \sum_{j=i+1}^N
 \,E^{\rm d-d}_{ij}
 \,\right\}\,.
 \label{totalE}
 \end{equation}
 The schematic view on its structure is displayed
 in Fig.\ref{Fig3}. According to the scheme of the computation,
 the interacting dots have their counterpart
 in the interacting RBFN blocks. This aspect makes
 our formulation close to the concept of the
 interacting neural networks \cite{Metzler2000}.

\section{The application to the array of rectangle
         ultrathin magnetic dots}\label{Square}

 In the concrete case we have modelled the ultrathin
 (quasi-two dimensional) rectangle magnetic dots
 of the square profile ${\ell}_{\rm d}\times {\ell}_{\rm d}$ and very
 small height $(V_{\rm d}/{\ell}_{\rm d}^2) $ $\ll {\ell}_{\rm d}$.
 Each dot is replaced by the four moments $N_{\rm c}=4$
 at the positions ${\bf r}_n=(1/4) 
 \ell_{\rm d}\,\,$
 $\Big[\,{\bf e}_x\,
 \cos\left(\frac{\pi}{2}(n-1)\right) $
 $+{\bf e}_y\,\sin\left(\frac{\pi}{2}(n-1)\right)\,\Big]\,$
 (see Fig.2(a)). The magnetic moment of ${\triangle}_{in}$
 element [see Fig.\ref{Fig2}(b)] maintains the saturation
 value $m_{\rm sat}= I_{\rm s} V_{\rm d}/4$.
 After this Eq.(\ref{lambdaeq})
 modifies to the form $\lambda = V_{\rm d}^2
 I_{\rm s}^2/(64\pi\mu_{\rm 0}\mu_{\rm r})$.

\begin{table}
\begin{center}
\caption{
The list of eleven memorized configurations of the magnetic
moments ${\tilde p}^{(q)}$ with the corresponding
self-energy parameters. The minus sign in ${\tilde p}^{(q)}$
substitutes~$-1$.
}
\vspace{5mm}
\begin{tabular}{|r|l|c|}
\hline
 &&
 \\
 $ q $
 &
 ${\tilde p}^{(q)}$
 &
 $ {\cal E}^{(q)} $
 \\
 \hline
 $  0  $
 &
 $ {\tilde p}^{(0)}\,\,=\Dot{\bullet}{\bullet}{\bullet}{\bullet}=
 \,\,\,\,\,\,\,(0,\,0,\,0,\,0,\,0,\,0,\,0,\,0\,)$
 &
 ${\cal E}_0$
 \\
 \hline
 $ 1   $
 &
 ${\tilde p}^{(1)}\,\,= \Dot{\uparrow}{\leftarrow}{\downarrow}{\rightarrow}
            = \kappa_{\rm v}\, (0,\,1,\,-,0,\,0,\,-,1,\,0)$
 &
 ${\cal E}_{\rm v}$
 \\
 $ 2 $
 &
 ${\tilde p}^{(2)}\,\,=\Dot{\downarrow}{\rightarrow}{\uparrow}{\leftarrow}
            = \kappa_{\rm v}\, (0,\,-,1,\,0,\,0,\,1,\,-,0)$
 &
 \\
 \hline
 $ 3 $
 &
 ${\tilde p}^{(3)}\,\,= \Dot{\uparrow}{\uparrow}{\uparrow}{\uparrow}
                       = \kappa_{\rm p}\, (0,\,1,\,0,\,1,\,0,\,1,\,0,\,1)$
 &
 ${\cal E}_{\rm p}$
 \\
 $ 4 $
 & ${\tilde p}^{(4)}\,\,= \Dot{\leftarrow}{\leftarrow}{\leftarrow}{\leftarrow}
                        = \kappa_{\rm p}\, (-,0,\,-,0,\,-,0,\,-,0)$
 &
 \\
 $ 5 $  & ${\tilde p}^{(5)}\,\,= \Dot{\downarrow}{\downarrow}{\downarrow}{\downarrow}
                        = \kappa_{\rm p}\, (0,\,-,0,\,-,0,\,-,0,\,-)$
 &
 \\
 $ 6 $  & ${\tilde p}^{(6)}\,\,= \Dot{\rightarrow}{\rightarrow}{\rightarrow}{\rightarrow}
                        = \kappa_{\rm p}\, (1,\,0,\,1,\,0,\,1,\,0,\,1,\,0)$
 &
 \\
 \hline
 $ 7 $ & ${\tilde p}^{(7)}\,\,= \Dot{\uparrow}{\leftarrow}{\uparrow}{\leftarrow}
          =\kappa_{\rm d}\, (0,\,1,\,-,0,\,0,\,1,\,-,0)$
 &
 ${\cal E}_{\rm d}$
 \\
 $ 8 $ & ${\tilde p}^{(8)}\,\,= \Dot{\downarrow}{\leftarrow}{\downarrow}{\leftarrow}
          =\kappa_{\rm d}\, (0,\,-,-,0,\,0,\,-,-,0)$
 &
 \\
 $ 9 $ & ${\tilde p}^{(9)}= \Dot{\downarrow}{\rightarrow}{\downarrow}{\rightarrow}
          =\kappa_{\rm d}\, (0,\,-,1,\,0,\,0,\,-,1,\,0)$
 &
 \\
 $ 10$ & ${\tilde p}^{(10)}\,\,= \Dot{\uparrow}{\rightarrow}{\uparrow}{\rightarrow}
          =\kappa_{\rm d}\, (0,\,1,\,1,\,0,\,0,\,1,\,1,\,0)$
 &
 \\
 \hline
 \end{tabular}
 \end{center}
 \end{table}

 We have proposed model including $Q=11$ memorized
 patterns (see Table~1) and four desired
 self-energy parameters
\begin{eqnarray}
 {w}^{(q)}=
\left\{
\begin{array}{lll}
{\cal E}_0        &  \mbox{\rm for}   &  q=0       \,,  \\
{\cal E}_{\rm v}  &  \mbox{\rm for}   &  q=1,2     \,,  \\
{\cal E}_{\rm p}  &  \mbox{\rm for}   &  q=3,4,5,6 \,,   \\
{\cal E}_{\rm d}  &  \mbox{\rm for}   &  q=7,8,9,10\,.
\end{array}
\right.
\label{eqqch}
\end{eqnarray}
 The additional parameters of the model are reduction 
 coefficients  
 $\kappa_{\rm v}\leq 1$,
 $\kappa_{\rm p}\leq 1$, $\kappa_{\rm d}\leq 1$
 (Table~1) introduced to modify 
 the size of memorized
 effective moments. 
 These coefficients describe the deviations 
 of moments from the saturated value. 
 According to Eq.(\ref{eqqch})
 one can introduce four subsets of vectors
 $\{  {\tilde p}^{(0)}  \}$,
 $\{  {\tilde p}^{(1)},  {\tilde p}^{(2)}  \}$,
 $\{  {\tilde p}^{(3)},  {\tilde p}^{(4)},
 {\tilde p}^{(5)}, {\tilde p}^{(6)}    \}$,
 $\{  {\tilde p}^{(7)},
 {\tilde p}^{(8)}, {\tilde p}^{(9)}, {\tilde p}^{(10)}   \}$.
 Within to each subset, the vectors correspond to the same 
 self-energy. 
 This system includes the
 vector pairs of the opposite sign
 $p^{(2)}=-p^{(1)}$, $p^{(5)}=-p^{(3)}$,
 $p^{(6)}=-p^{(4)}$,  $p^{(9)}=-p^{(7)}$,  $p^{(10)}=-p^{(8)}$
 and zero memorized vector ${\tilde p}^{(0)}$. 
 This structure
 guarantee the reflection symmetry
 of the self-energy given by Eq.(\ref{invariance1}).

 The exceptional vector ${\tilde p}^{(0)}$ concerns
 the integral information from multidomain
 or chaotic magnetization modes of the oscillatory
 or chaotic character \cite{Chou97}. Its occurrence
 is a signature of uncertainty in description of a high
 momentum magnetization modes. Among the patterns
 memorized and restored by RBFN, we focussed attention
 to the vortex magnetization modes
 \cite{Chou97,Vadmedenko99,Shinjo2000,Fernandez98}.
 The next two vectors ${\tilde p}^{(1)}$ ,${\tilde p}^{(2)}$
 encode the symmetric vortex and counter-vortex configurations
 in Fig.\ref{Fig2}(c), Table~1. Similarly, as in the case
 of ${\tilde p}^{(0)}$, the total magnetic moment of the
 symmetric memorized vortex is zero. For the rectangular
 ultrathin isolated dots of square profile and small
 crystalline anisotropy, the vortex type of magnetic
 ordering was revealed by the Monte-Carlo simulations
 \cite{Vadmedenko99}. This finding was confirmed
 by the experiments \cite{Shinjo2000,Fernandez98}.
 Vortex modes were also detected by the simulations
 on a cubic particles \cite{Bertram88} for a weak or zero external
 magnetic fields. The system
 of vectors $\{ {\tilde p}^{(3)}$,  ${\tilde p}^{(4)}$,
 ${\tilde p}^{(5)}$, ${\tilde p}^{(6)} \}$
 belonging to the Stoner-Wohlfart type of single-domain
 particle \cite{Brown62} is represented by the four parallel
 effective moments. This ordering can also occur by virtue
 of the external magnetic or magnetostatic fields. The remaining
 intradot configurations labeled by $q=7,8,9,10$
 should be the potential sources of (shape) anisotropy.
 Let us to note that RBFN approach is not sensitive
 to the physical nature of the anisotropy.

 The principal question arises how to determine
 seven single-dot parameters $\kappa_{\rm v}$,
 $\kappa_{\rm p}$,
 $\kappa_{\rm d}$,
 ${\cal E}_0$,
 ${\cal E}_{\rm v}$,
 ${\cal E}_{\rm p}$
 and ${\cal E}_{\rm d}$. Further work is needed
 to combine the present simulations with the
 micromagnetic approaches
 (see e.g.\cite{Popkov99,Scheinfein90})
 incorporating the algorithms of the neural network
 learning \cite{Haykin98}. In this paper the magnetic
 configurations were selected and parametrized
 in heuristic manner.

\section{The implementation of simulated annealing method}\label{Simull}

Simulated annealing \cite{Kirkpatrick83} is an
optimization technique which operates in a manner analogous
to the physical process of annealing. In this section
we discuss some details of its implementation to dot
array model. The subject of minimization is energy
functional Eq.(\ref{totalE}) of the effective magnetic moments.
The main parts of adopted algorithm are:
\begin{itemize}
\item[\bf 1.]  {\em Initial state}
               ${\tilde m}_i(t=0)$ is generated
               (or read from the data file).
\item[\bf 2.]  {\em Cooling schedule}. The pseudotemperature
               $T(t)= T_0\,\exp\left(\,-t/t_0 )\right)$
               relaxes as a function of the discrete
               time $t=0,1,\ldots, t_{\rm max}$.

\item[\bf 3.]  {\em The update equation} is based on the standard
           algorithm of Metropolis \cite{Binder92,Metropolis53}.

\item[\bf 3.1] {\em Single moment moves}.
               The lattice dot index
               $i\in \{1,2,\ldots, N\}$, and interaction center
               index $n\in \{1,2,\ldots, N_{\rm c}\}$ are chosen randomly.
               At given time $t$, the current effective moment
               ${\bf m}_{in}(t)$ undergoes the
           stochastic single-moment move
           $ {\bf m}_{in}^{\small\rm trial}(t+1)=
           {\bf m}_{in}(t) + v_{\rm m}\, {\bf Y}(t)\,,$
           where ${\bf Y}$ is two-dimensional vector
          generated over uniform distribution
          in $<-1,1>~\times~<-1,1>$. The generation is repeated
          until the constrain
          $|{\bf m}_{in}(t) + v_{\rm m}\, {\bf Y}(t)|\leq 1$
          is satisfied;
          $v_{\rm m}$ is the parameter,
          which controls the remagnetization speed
          $\left| {\bf m}(t+1) - {\bf m}(t) \right|
          \leq v_{\rm m}\,/(N\,N_{\rm c}) $.
          The encoding
          of ${\bf m}^{\small\rm trial}_{in}(t)$
             gives rise to the row vector
             ${\tilde m}_i^{\rm trial}$(t).
             By taking into
             account Eq.(\ref{totalE})
             with the concrete form of the Nadaraya-Watson
             self-energy estimator [Eq.(\ref{EqNW1})]
             and expression for dipole-dipole
             interaction Eq.(\ref{dipolar})
             we obtained for the energy
             variation $\Delta E_{in}$
             associated with the elementary move
             from ${\bf m}_{in}(t)$ to
             ${\bf m}_{in}^{\small\rm trial}(t+1)$ as
\begin{eqnarray}
&&\Delta E_{in}=\sum_{q=0}^{Q-1}
{w}^{(q)}\,
\left[\,\,
\psi^{(q)}({\tilde m}_i^{\rm trial})-
\psi^{(q)}({\tilde m}_i)\,\,
\right]
 \label{deltanerg}
 \\
 &&-
\lambda
\sum_{\alpha=x,y}
\left(
\,\,m_{in,\alpha}^{\small\rm trial}- m_{in,\alpha}
\,\,\right)
\left[\,\,
h_{\alpha} -
\sum_{\beta=x,y}\,\,
\sum_{j=1,j\neq i}^{N}\,
\sum_{s=1}^{N_{\rm c}}\,
A_{injs,\alpha\beta}\,
m_{js,\beta}\,\,
\right]
\nonumber
\end{eqnarray}

 \item[-]  {\em Acceptance criteria}
           If $\Delta E_{in}\leq 0$,
           the trial configuration
           is accepted automatically
           and
           ${\bf m}_{in}(t+1)={\bf m}_{in}^{\rm trial}(t+1)$.
           If $\Delta E_{in}>0$,
           the trial configuration is accepted
           if the Boltzmann factor
           $\exp(-\Delta E_{in}/T)$ is larger
           or equal to the random number
           generated uniformly
           over $<0,1>$.

 \item[-]   {\em Quasistatic simulations}.
            In the zero temperature limit,
            the acceptance criteria reduces
            to the absolute
            acceptance if $\Delta E_{in}\leq 0$. For this
            dynamics the energy is monotonically decreasing
            function of time. Subsequently, the stochastic
            motion through the phase space tends to the accessible
            basin of attraction. In the matastable state
            the motion gets stuck for a fixed external magnetic
            field. The sequence of the metastable configurations
            obtained for a gradually changing external
            magnetic field was used for the calculation
            of the quasistatic hysteresis loops.

 \item[{\bf 3.2}] {\em Complex intradot moves} allows
            to enhance the effectiveness of annealing
            and left the metastable states.
            The update according to ${\bf 3.1}$ is
            supplemented~by~:

 \begin{itemize}

 \item[\bf 3.2a] {\em Interchange moves} changing
             the location of two intradot configurations:
             ${\tilde m}_i^{\rm trial}={\tilde m}_j$,
             ${\tilde m}_j^{\rm trial}={\tilde m}_i$
             belonging to two randomly chosen dots
             $i$ and $j$.

 \item[\bf 3.2b] {\em Feature moves} starting with
             the random choice of the dot index $i$
             and its feature
             $q\in\Lambda_Q$,
             $q \neq {\cal F}({\tilde m}_{i})$.
             The suggested move
             is then given by
             ${\tilde m}_i^{\rm trial} = {\tilde p}^{(q)}$.

 \item[\bf 3.2c] {\em Reflection moves} defined by
             ${\tilde m}_i^{\rm trial}= -{\tilde m}_i$
             are of the special importance since they
             conserve the self-energy
             [see Eq.(\ref{invariance1})].
             Due to this property the simulated system 
             can overcome
             easily the self-energy 
             barriers.

 \end{itemize}

 For the moves ${\bf 3.2 a,b,c}$
 the acceptance
 probability is given by

 $$\min\left\{\,1,\exp\left(-\sum_{n=1}^{N_{\rm c}}\Delta
 E_{in}\right)\,\right\}.$$

 \item[\bf 4.] {\em Stopping criteria}.
           For $t<t_{\rm max}$,
           the annealing  process
           follows from the step ${\bf 2}$ with $t\leftarrow t+1$
           for the move ${\bf 3.1}$, or with
           $t\leftarrow t+N_{\rm c}$
           in the case of the
           complex intradot moves
           ${\bf 3.2~a,b,c}$.

\end{itemize}

\section{Numerical simulations}\label{NumSimull}

\begin{figure}
\begin{center}
\caption{
The system of dot centers ${\bf R}_1, {\bf R}_2$,
$\ldots, {\bf R}_{193}$ of the simulated cluster.
The unit vectors ${\bf e}_x$ and ${\bf e}_y$
are parallel to the main directions of the
square lattice.}
\vspace*{3mm}
\includegraphics[width=9.6cm,height=6.4cm]{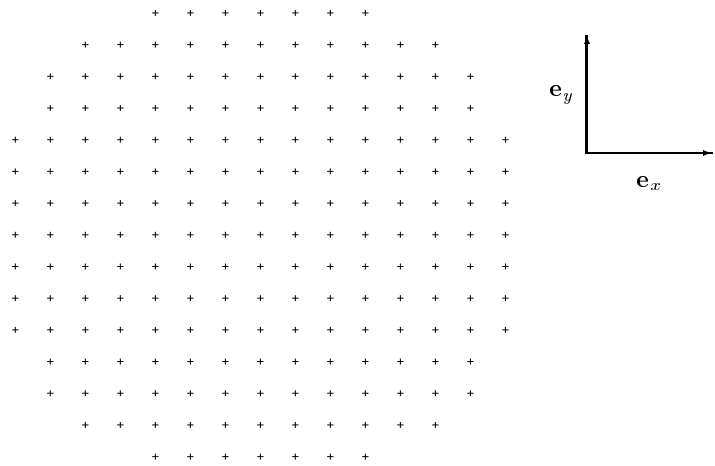}
\vspace{3mm}
\label{Fig1}
\end{center}
\end{figure}

\begin{figure}
\caption{The basic parameters of the dot array model:
(a)~$i$th dot is located at the position ${\bf R}_i$;
the magnetic structure of the dot is represented by the four
interaction centers and four effective magnetic moments
${\bf m}_{in}$; (b)~two examples of the dots showing how
the effective magnetic moment arises from the integration
of the magnetization field over the $\triangle_{in}$ element;
(c)~the example of the memorized intradot configuration consisting
of the four magnetic moments. The example of the encoding
of intradot
configuration by means of the eight-dimensional
vector ${\bf p}^{(1)}$.}

\vspace{4mm}
\begin{center}
\includegraphics[width=8.0cm,height=14.0cm]{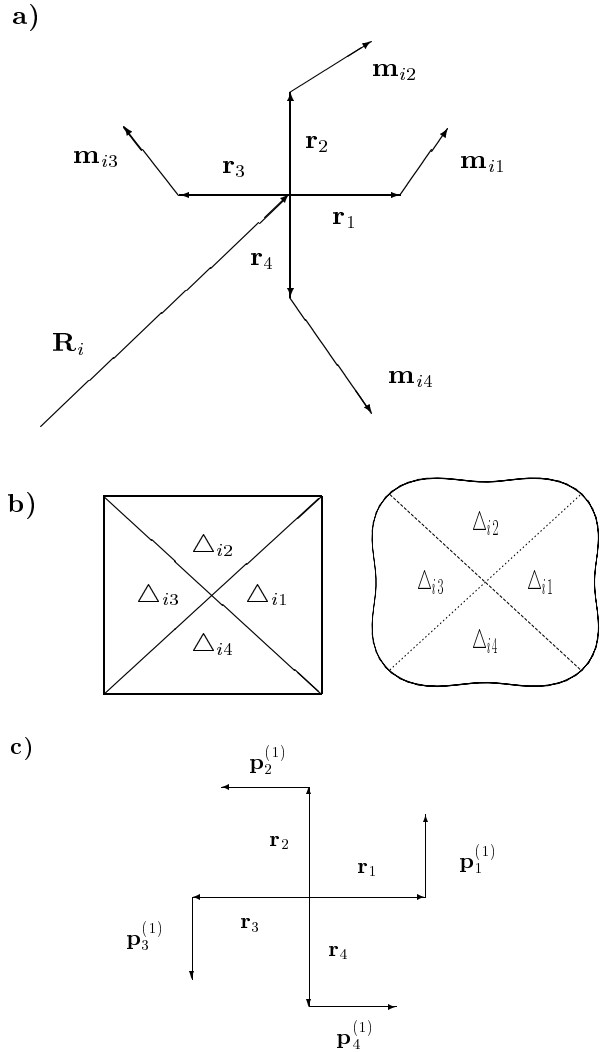}
\end{center}
\vspace{4mm}
\label{Fig2}
\end{figure}

 For the model defined in Sect.\ref{Square} we performed
 the numerical simulations. We studied finite disk-shaped
 cluster $R^{\rm cluster}=8\,a$ including $N=193$ dots
 of the size $l_{\rm d}=a/8$. We assume that each dot
 has $N_{\rm c}=4$ centers at the square lattice
 [see Fig.\ref{Fig1}]. The parameters of memorized
 magnetic configurations are
 $\kappa_{\rm v}=\kappa_{\rm d}=0.95$,
 $\kappa_{\rm p}=1$. Consequently, $d_{\rm max}=4$.
 The typical parameters of the simulated annealing
 have been
 $v_{\rm m}=0.2$,
 $t_0 = 200 N N_{\rm c}$,
 $t_{\rm max} =  5 t_0$,
 and $ \lambda = T_0 \leq 10  \lambda$.
 For the quasistatic simulations  we used
 $t_{\rm max}=100 N N_{\rm c}$.

 The initial simulations were performed for a zero
 external field and zero self-energy parameters.
 In this case the Monte-Carlo minimization
 of the energy leads to the rapid fall-off
 of the energy towards the non-colinear
 antiferromagnetic chains. The configuration
 is displayed in Fig.\ref{Fig8}(a). From this follows
 that central part of this cluster corresponds
 to the noncolinear antiferromagnetic phase
 in agreement with magnetic configuration
 obtained for a system of cylindrical dots
 \cite{Gushlienko99} and truncated
 dipolar moments \cite{Prakash90}. 
 At the same time
 the surface moments 
 which tend to be parallel to the cluster surface
 exhibit some kind of the frustration
 \cite{Vadmedenko99}. The annealing leads to the
 ground state estimate $E=E^{\rm AF}=-38.77 \lambda N$.

 The previous value can be understood as the threshold
 for the competition between the interdot and intradot
 structures. The natural way of the stabilization
 of the intradot vortices is to make the parallel
 structures of the moments $(q=3,4,5,6)$ energetically
 unfavorable. In the next we will analyzed more restrictive
 choice: $w^{(q\neq 1,2)} \geq | E^{\rm AF}|$ with
 the calibration condition $w^{(q=1,2)}={\cal E}_{\rm v}=0$.

\begin{table}
\caption{ The numerical tests of Nadaraya-Watson formula calculated
for two sets of the self-energy parameters. The comparison of
desired and retrieved energies [see Eq.(\protect\ref{condition1})].
The formula exhibits weakly non-uniform
response $E^{\rm self}({\tilde p}^{(q\geq 1)})$
to the uniform input
${\cal E}^{(q\geq 1)}=60\lambda$.
}
\begin{center}
\begin{tabular}{|l|r|r||r|r|}
\hline
$q$                                        &
${\cal E}^{(q)}/\lambda$                    &
$E^{\rm self}({\tilde p}^{(q)})/\lambda$   &
${\cal E}^{(q)}/\lambda$                   &
$E^{\rm self}({\tilde p}^{(q)})/\lambda$
\\
\hline
0             &    20   &   27.72  &     0  &   25.85    \\
1, 2          &     0   &    2.84  &    60  &   56.50    \\
3, 4, 5, 6    &    60   &   55.02  &    60  &   56.88    \\
7, 8, 9, 10   &    40   &   40.71  &    60  &   56.82    \\
\hline
\end{tabular}
\end{center}
\end{table}

\begin{figure}
\caption{
  The hysteresis loops of the isolated
  dots simulated  for RBFN parameters:
a)~see Eq.(\ref{casea});
b)~$ {\cal E}_{\rm v}=60\lambda ,
     {\cal E}_{\rm p}=0         ;
     {\cal E}_{\rm d}=60 \lambda,
     {\cal E}_0=60 \lambda      ; $
c)~$ {\cal E}_{\rm v}=60\lambda ,
     {\cal E}_{\rm p}=20\lambda ,
     {\cal E}_{\rm d}=0         ,
     {\cal E}_0=60\lambda       ; $
d)~$ {\cal E}_{\rm v}=0         ,
     {\cal E}_{\rm p}=60 \lambda,
     {\cal E}_{\rm d}=60 \lambda,
     {\cal E}_0=0\,.              $
}
\vspace{5mm}
\begin{center}
\includegraphics[width=13.0cm,height=10.0cm]{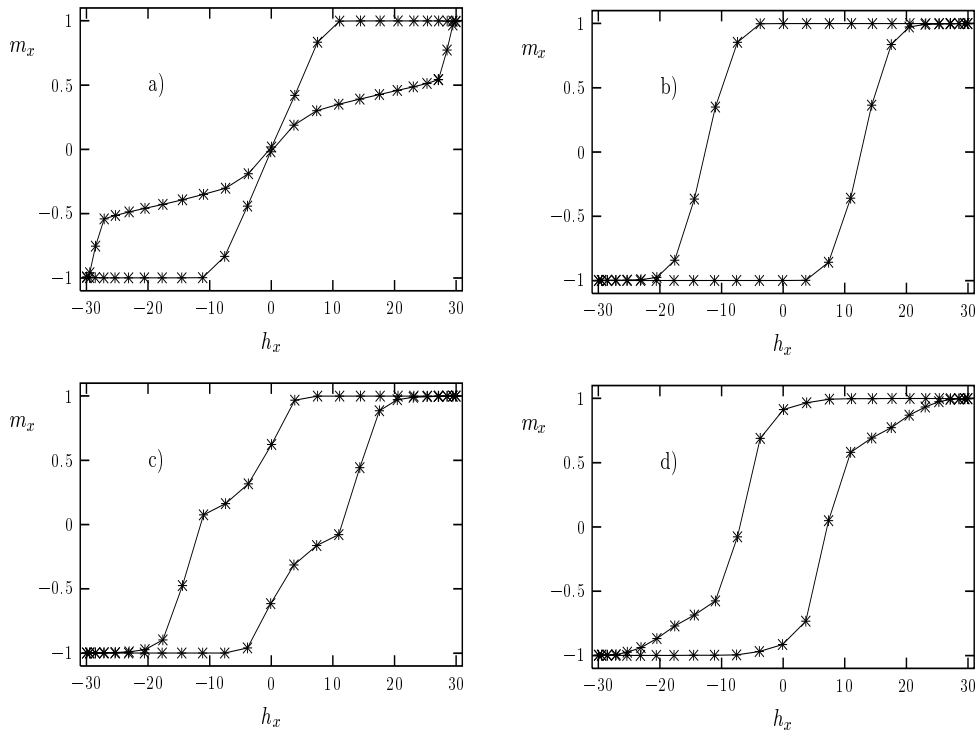}
\end{center}
\vspace{5mm}
\label{Fig4}
\end{figure}

\begin{figure}
\caption{The hysteresis loop of the interacting
         dots in the anisotropic case, calculated
         for $\theta=0^o$ and paramters
         from Eq.(\ref{casea}).
}
\vspace{5mm}
\begin{center}
\includegraphics[width=12.0cm,height=6.0cm]{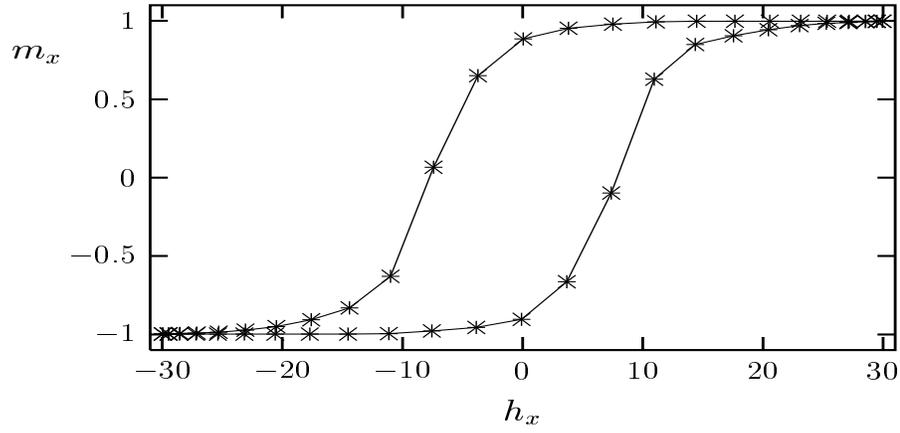}
\end{center}
\label{Fig5}
\vspace{5mm}
\end{figure}

 The previously simulated system was purely magnetostatic.
 We follow with the simulations of the opposite kind of
 systems, where interdot interactions have been completely
 neglected. For these systems we have constructed
 a quasi-static hysteresis loops. The results
 have been obtained for the several combinations
 of the self-energy
 parameters (comparable
 with $| E^{\rm AF}|$).
 They are presented in Figs.\ref{Fig4}(a)-(d).
 In Table~2 we list differences between the desired
 self-energies and outputs of Nadaraya-Watson estimator.
 Because our parameters are free, the inaccuracy stemming
 from Nadaraya-Watson formula has no principal
 significance for the quality of the result.
 In the situations, where precision of output
 becomes to be more relevant, the sofisticated
 RBFN learning is required \cite{Haykin98}.
 The interesting situation has occurred for the self-energy
 parameters
 \begin{eqnarray}
 {\cal E}_0 =      20  \lambda\,,
 \qquad
 {\cal E}_{\rm v}= 0   \,,
 \qquad
 {\cal E}_{\rm p}= 60  \lambda\,,
 \qquad
 {\cal E}_{\rm d}= 40  \lambda\,
 \label{casea}
 \end{eqnarray}
 corresponding to Fig.\ref{Fig4}(a). The field dependence
 of ${\overline n}(q)$ observed during the remagnetization
 process for $q=1,2$ confirms the vortex stabilization
 around $h\simeq 0$ and zero remanence. Qualitativelly similar
 behavior with the vortex annihilation and formation was observed
 in the experimental study \cite{Fernandez98}. The situation changes
 dramatically when the strong interdot interactions are taken
 into account. Fig.\ref{Fig5} shows that their influence
 causes the non-zero ramanence due to suppression of vortices:
 ${\overline n}(q)\leq 0.07$ for $q=1,2$. More detailed
 analysis of this fact has revealed that surface vortices
 are more stable than vortices from the central zones of the
 cluster. Two next figures \ref{Fig6}(a),(b) show how the
 anisotropy of the memorized configurations is reflected by
 the hysteresis loops of magnetization components
 $m_x(h)$, $m_y(h)$ constructed for $\theta=20^o$.

\begin{table}
\caption{ The comparing of the several numerical results
obtained the external magnetic fields
$h=0,\, 5,\, 10$. Table shows structure of energy contributions
(divided by $N\lambda$) to the ground state.
Calculated for the
parameters from Eq.(\ref{casea}).
}
\begin{center}
\begin{tabular}{|l||r|r|r|}
\hline
energy                                    &
$h=0 $                                    &
$h=5 $                                    &
$h=10$
\\
\hline
total             &    1.58   &    -3.41   &    -11.85      \\
magnetic field    &    0.00   &    -8.25   &    -23.10      \\
self              &    6.37   &    13.50   &     24.23      \\
magnetostatic     &   -4.78   &    -8.67   &    -12.98      \\
\hline
\end{tabular}
\end{center}
\end{table}

 In the case with the non-zero self-energy parameters
 interact magnetostatically, the choice of the initial
 conditions of the simulated annealing becomes
 to be more complicated. Several final configurations
 obtained by the annealing process are displayed in
 Figs.\ref{Fig8}(b)-(i). The preliminary runs evolving
 from the initial random state get stuck in the local
 minimum $E=6.4\lambda N$. The configuration of this
 metastable state is displayed in Fig.\ref{Fig8}(b).
 The application of the feature map [see Fig.\ref{Fig8}(c)]
 reveals that the clustering of the dot states resembles
 the formation of the homogeneous domains in $Q-$state
 Potts model \cite{Mouritsen89}. The
 lowest energy $E=1.58 \lambda N$ was obtained for the system
 initialized from the vortex state. No essential
 differences between the pure vortex and mixed
 vortex-countervortex initial conditions
 were observed, contrary to our expectation evoked
 by the study of dipolar system \cite{Vadmedenko99}. The question of
 the helicity will require more detailed investigation.
 Fig.\ref{Fig8}(d) shows that magnetostatic deformation
 of vortices is rather pronounced feature. After
 the deformation, four-moment vortices acquire the
 nonzero magnetic moments and resemble the fans
 or vortices with the non-central N\'eel-type core.
 Their analysis via ${\cal F}$ map shows 
 the mixing
 of vortex 
 features $q=1,2$ 
 [ $ {\overline n}(1) $ $ \simeq $ ${\overline n}(2)$  
 $\simeq 0.5 $]
 and separation of the clusters with different vorticity
 [see Fig.\ref{Fig8}(d). This aposteriory finding
 confirms that parameters of Eq.(\ref{casea}) are
 sufficient for the stabilization of the vortex
 ground state for $h=0$. Two configurations displayed
 in Figs.\ref{Fig8}(e),(f) were obtained for
 the external magnetic field. Their energies
 are listed in Table~3. For $h=h_x=5$ the
 field-deformed vortices resemble the fans ordered into
 the large-scale wave-like structures. The waves
 are better visible from Fig.\ref{Fig8}(h) showing the detailed snapshot,
 where intradot magnetic moments are averaged
 for each dot separatelly. The chaining of field-oriented
 phase is also visible. For $h=h_x=10$ Fig.\ref{Fig8}(i)
 demonstrates the formation of the clusters with
 $q=6$.

 To characterize the anisotropy, we have
 studied the angular dependence of the magnetization
 for $h=5$ and varying $\theta\in<0^o,180^o>$.
 In the simulations we have distinguished between
 the clockwise and counter-clockwise field rotation directions.
 After the annealing starting from the purely vortex
 state ${\tilde m}_{1\leq i \leq N}(t=0)={\tilde p}^{(1)}$,
 $\theta=0^o$ ($\theta=180^o$)
 we performed the series
 of quasistatic remagnetization steps
 at zero temperature.
 These simulations have revealed the hard
 axes ${\bf e}_x$, ${\bf e}_y$ and easy
 axes ${\bf e}_x\pm {\bf e}_y$
 [see Fig.\ref{Fig7}]. In addition,
 the model system
 exhibits the angular hysteresis. These results
 demonstrate how the outputs of RBFN
 mimic the biaxial
 anisotropy and how the anisotropy
 is reinforced by the magnetostatic
 couplings.

 \section{Conclusions}

We believe that very general method we have
introduced in this paper will be stimulating
for the people working in the field
of the micromagnetic simulations of the nanoscale systems.
From the point of view of magnetostatics, direct model
improvement is possible in many ways:
a)~near dot interactions can be taken into account more accuratelly
   by including the rectangle-rectangle magnetostatic terms;
b)~to speed-up the computations and to extend the system size
   one can use the hierarchical summation \cite{Miles91}.
We thing that more realistic simulations will
be possible after the finding of a closer relationship between
the neural networks and outputs of the standard
micromagnetic approaches.

\noindent {\bf Acknowledgement}: This work
was supported by the grant no.1/6020/99 and by
the Polish-Slovak international grant.

\newpage
\section{Figure Captions}


\begin{figure}
\caption{The hysteresis loops of interacting
dots in the anisotropic case for $\theta=20^o$.
Calculated performed for the parameters taken from Eq.(\ref{casea}).}
\vspace{5mm}
\begin{center}
\includegraphics[width=13.0cm,height=5.0cm]{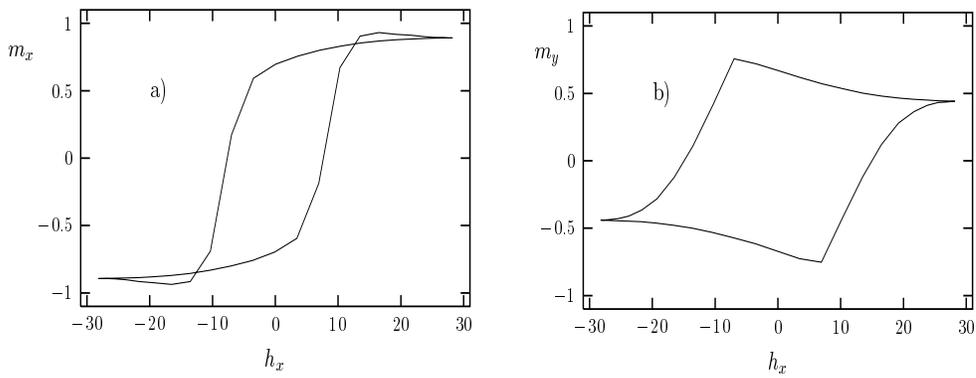}
 \end{center}
\vspace{5mm}
\label{Fig6}
\end{figure}


\begin{figure}
\caption{The angular dependence of
the magnetization for interacting
(int.) and non-interacting (no int.)
dot systems confirming the biaxial anisotropy.
The arrows indicate direction of the field
rotation. Calculated for $h=5$ and
parameters Eq.(\ref{casea}).
}
\vspace{5mm}
\begin{center}
\includegraphics[width=10.0cm,height=7.0cm]{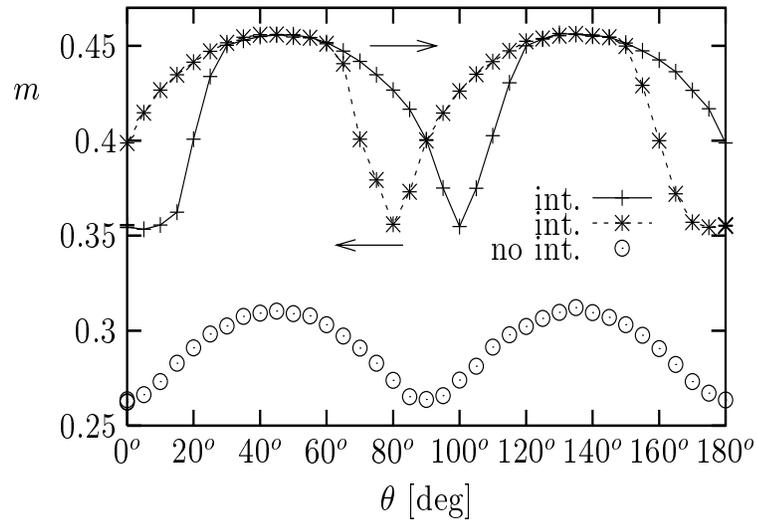}
\end{center}
\vspace{5mm}
\label{Fig7}
\end{figure}


\begin{figure}
 \begin{center}
\caption{
 The system of the low energy configurations revealed by the simulated
 annealing process.
 The case~(a) belongs to the zero self-energy
 parameters and essentially magnetostatic
 antiferromagnetic configuration obtained for
 $h=0$. The configuration shows only the moments
 obtained for each dot separatelly. The annealing
 provides $E=E^{\rm AF}=-38.77\lambda N$.
 The results obtained in cases (b)-(i) belong
 to the parameters from Eq.(\ref{casea}).
 Part~(b) displays disordered - quenched metastable
 microstate of the energy $E=6.4\lambda N$ (for $h=0$)
 obtained for the random initial condition
 and forbiden moves {\bf 3.2~(a),(b),(c)}. The metadomain
 structure was analyzed by ${\cal F}$ map [case (c)].
 The annealing from the initial vortex state stabilizes
 at the lowest energy $E=1.58 \lambda N$ [cases (d),(g)].
 Part~(g) shows only the dots with $q=1$,
 the remaining part with $q=2$ was removed for the
 clearness. For the non-zero external fields we constructed
 snapshots (e),~(h)~$h_x=5$,  $h_y=0$ (wave-like ordered
 structures); (f),~(i)~$h_x=10$, $h_y=0$                                 ;
 Part~(h) shows averaged magnetic moments per each dot.
 It forms wave-like structure. For~(i) the formation
 of $q=6$ phase clusters is visible
 (here labels of $q=1,2$ features are 
 removed).
 }
 \vspace{5mm}
 \includegraphics[width=13.0cm,height=15.8cm]{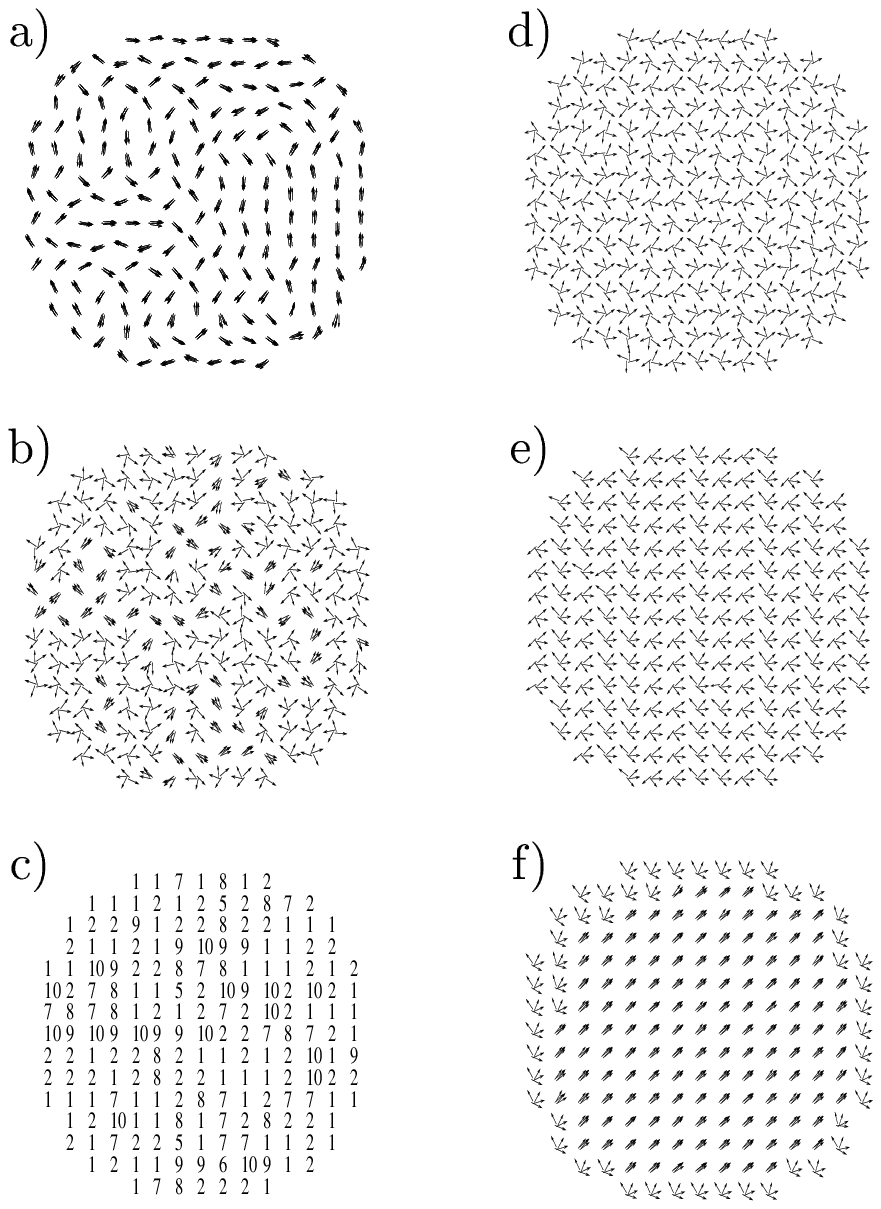}
 \vspace{5mm}
 \label{Fig8}
 \end{center}
 \end{figure}

 \end{document}